\documentstyle{mn}

\newcommand{\etal}{{et al.}}

\newcommand{\Ovii}{\mbox{{O{\small VII}}}}
\newcommand{\Oviii}{\mbox{{O{\small VIII}}}}
\newcommand{\Neix}{\mbox{{Ne{\small IX}}}}
\newcommand{\Nex}{\mbox{{Ne{\small X}}}}
\newcommand{\Nexi}{\mbox{{Ne{\small XI}}}}

\newcommand{\Mgxi}{\mbox{{Mg{\small XI}}}}
\newcommand{\Mgxii}{\mbox{{Mg{\small XII}}}}

\newcommand{\Sixiii}{\mbox{{Si{\small XIII}}}}

\newcommand{\Fexvii}{\mbox{{Fe{\small XVII}}}}
\newcommand{\Fexviii}{\mbox{{Fe{\small XVIII}}}}
\newcommand{\Fexix}{\mbox{{Fe{\small XIX}}}}
\newcommand{\Fexx}{\mbox{{Fe{\small XX}}}}

\newcommand{\Fexxiv}{\mbox{{Fe{\small XXIV}}}}

\newcommand{\gtsim}{\raisebox{-1mm}{$\stackrel{>}{\sim}$}}
\def\arcm{\hbox{$^\prime$}}
\def\arcs{\arcm\hskip -0.1em\arcm}

\title[{\em XMM-Newton} RGS Observations of M82]{{\em XMM-Newton}
Reflection Grating Spectrometer Observations of the Prototypical
Starburst Galaxy M82}

\author[A.M. Read \& I.R. Stevens]{Andrew
M. Read$^{1}$ \& Ian R. Stevens$^{1}$\\  
$^{1}$ School of Physics and Astronomy, University of Birmingham, 
Edgbaston, Birmingham, B15 2TT, UK\\
(E-mail: amr@star.sr.bham.ac.uk, irs@star.sr.bham.ac.uk)} 

\date{Accepted ..............................; 
Received ..............................; 
in original form ..............................}

\begin{document}

\maketitle

\begin{abstract} 

We present results from {\em XMM-Newton} Reflection Grating
Spectrometer observations of the prototypical starburst galaxy
M82. These high resolution spectra represent the best X-ray spectra to
date of a starburst galaxy. A complex array of lines from species over
a wide range of temperatures is seen, the most prominent being due to
Lyman-$\alpha$ emission from abundant low Z elements such as N, O, Ne,
Mg and Si. Emission lines from Helium-like charge states of the same
elements are also seen in emission, as are strong lines from the
entire Fe$-$L series. Further, the \Ovii\ line complex is resolved and
is seen to be consistent with gas in collisional ionization
equilibrium.

Spectral fitting indicates emission from a large mass of gas with a
differential emission measure over a range of temperatures
(from $\sim$0.2\,keV to $\sim$1.6\,keV, peaking at $\sim$0.7\,keV), and
evidence for super-solar abundances of several elements is indicated.
Spatial analysis of the data indicates that low energy emission is
more extended to the south and east of the nucleus than to the north
and west. Higher energy emission is far more centrally concentrated.

\end{abstract}


\section{Introduction}

Starbursts play a key role in galaxy formation and evolution,
and, in addition, feedback from star-formation in starburst galaxies, in
the form of superwinds, probably plays a key role in enriching and
heating the intergalactic medium (see Heckman 2001 for a recent
review). Further, because superwinds are driven by hot gas generated by
stellar winds and supernovae in the starburst, X-ray observations of
starbursts have a pivotal position in understanding the structure of
starburst and superwinds.

M82 is regarded as one of the archetypical starburst and superwind
galaxies. It is a nearby object ($D=3.63$\,Mpc, Freedman \etal\ 1994),
is very IR luminous, and is currently undergoing a strong starburst
(Rieke \etal\ 1993). M82 has been observed by all major X-ray
satellites (e.g.\,Read, Ponman \& Strickland 1997; Strickland, Ponman
\& Stevens 1997; Cappi \etal\ 1999; Kaaret \etal\ 2001), and the
superwind is clearly visible in X-rays and also in H$\alpha$ (Shopbell
\& Hawthorn 1998), with clear evidence of a bipolar outflow and a good
deal of similarity in the X-ray/H$\alpha$ morphology (Lehnert, Heckman
\& Weaver 1999). Devine \& Bally (1999) noted a feature termed the
H$\alpha$ \lq\lq cap\rq\rq\ lying well above the galaxy disc 
that has X-ray emission associated with it.

The {\em XMM-Newton} satellite has the largest collecting area of any
imaging X-ray telescope, and this coupled with the spectral-imaging
(EPIC) and grating instruments (Reflection Grating Spectrometers $-$
RGS) offers a new window on starburst galaxies. As M82 is the
brightest starburst in the sky, these observations are of key
importance to understanding starbursts. In this paper we present
results from the RGS. Preliminary results from the 
EPIC instruments will be presented in Bravo-Guerrero \etal\ (2002).
It is worth noting that at the present time, almost no RGS results
from non-active spiral galaxies have been presented, the most relevant
being the very preliminary results from the other famous nearby
starburst, NGC~253 (Pietsch \etal\ 2001).

The RGS instruments (den Herder \etal\ 2001) on board the European
X-ray observatory {\em XMM-Newton} (Jansen \etal\ 2001) offer, at
least for objects of extent of order 1$-$2\arcm, the
possibility of very high resolution X-ray spectroscopy combined with
some spatial resolution in the cross-dispersion direction. Such unique
capabilities are excellently suited to the study of the thermodynamic
properties of the hot gas within and surrounding the M82 starburst
nucleus. Here we describe our preliminary findings in
analysing the RGS data obtained during the {\em XMM-Newton} guaranteed
time observations. Results from further more in-depth analysis will be
presented in a future paper. The next section describes the
observations and the data reduction techniques. Preliminary
spectral and spatial results are then presented and discussed, and
finally we present our conclusions.

\section{Observations and data reduction}

M82 was observed with {\em XMM-Newton} during orbit 258 on the 5th of
June 2001 for one exposure of around 30\,ks for each of the European
Photon Imaging Camera (EPIC) instruments and the RGS spectrometers.
The standard reduction of the RGS data was performed using the latest
version (V.\,5.2) of the Science Analysis System (SAS), using the most
recent calibration for effective area and wavelength scale. The
reduction was performed using {\em rgsproc-0.98.4}, defining the
position of the M82 nucleus implicitly.

M82 is a very bright X-ray source, and thus we were not too concerned
with variations in the background. Nevertheless, a very short period
of high background proton flaring was removed from the data. In
extracting the spectra, we attempted to optimize the location and
width of the spatial extraction mask. Such a process can involve a
trade-off between improving the signal-to-noise via maximizing the
number of source counts, and reducing the background noise via
minimizing the source extraction region. The fraction of the PSF
sampled within the default pipeline mask is 90 per cent of the source
counts. This assumes however, that the source is point like. In our
analysis, we used a value of 95 per cent. The width of the resultant mask in
the cross-dispersion direction ranged from 1.3$-$2\arcm, and
covered the vast majority of the emission visible in the RGS. The
background within this mask was seen to be largely negligible.

\begin{figure*}
\vspace{10.4cm} 
\includegraphics{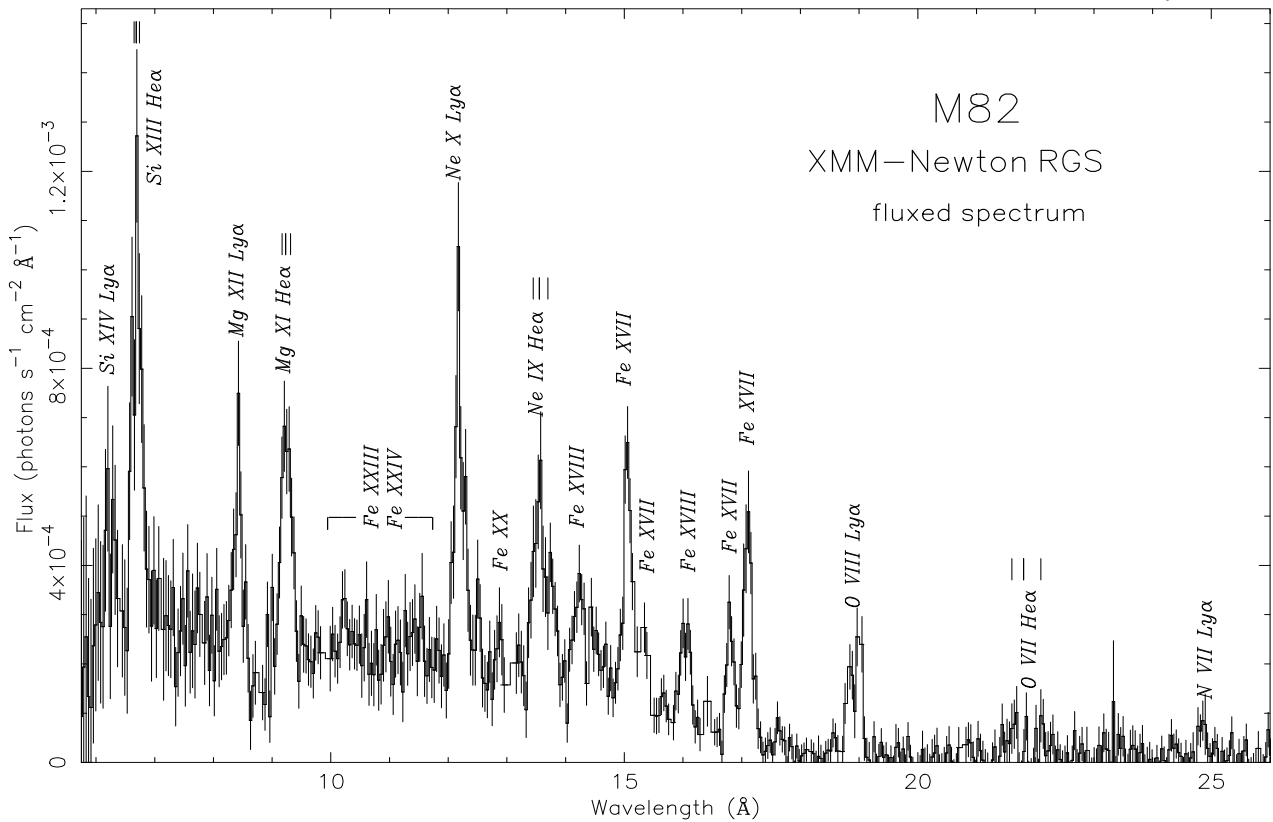} 
\caption{The fluxed RGS spectrum (RGS 1 \& 2, order 1) in counts
cm$^{-2}$ s$^{-1}$ \AA$^{-1}$ of the nuclear M82 starburst. The
identifications of bright emission lines are indicated (see text for
details). Resonance-intercombination-forbidden triples of the He-like
species are also indicated (vertical lines).}
\end{figure*}

As the spatial resolution of the RGS in the cross-dispersion direction
is about 15\arcs, 5 to 8 spatial resolution elements are
sampled within our extraction region. The {\em XMM-Newton} EPIC images
(Bravo-Guerrero \etal\ 2002) show the bright M82 core to have an
extent of $\sim$15\arcs, and we expect the RGS lines to be
broadened. The broadening varies only a little with wavelength and is
approximately $0.124\theta$\,\AA, $\theta$ being the source extent
(in arcminutes). Hence the broadening should be about 0.03\AA, and we
should be able to resolve the helium-like triplets of \Ovii, and possibly
\Neix, though perhaps not those of \Mgxi\ and \Sixiii.

First- and second-order spectra for RGS1 and RGS2 were
extracted. While it is possible and tempting to merge the data into
one spectrum, this is not encouraged for analysis purposes as the
pointing of the two RGS instruments are different, resulting in
different dispersion angles and wavelength scales. 
It is useful however to combine the spectra for non-analysis purposes
to obtain a single spectrum, and increase the signal-to-noise to
detect faint spectral features. A single fluxed spectrum, in units of photons
cm$^{-2}$ s$^{-1} \AA^{-1}$, was produced from the two first-order spectra and
is shown in Fig.~1 (second-order spectra can be added also, though this
degrades the spectral resolution).

As can be seen, the RGS spectrum is dominated by Ly$\alpha$ emission
lines of hydrogenic charge states of the abundant low Z elements (N,
O, Ne, Mg, Si). Helium-like charge states of the same elements are
also seen in emission, growing in strength (relative to the respective
Ly$\alpha$ emission) with line energy (from O to Si). Neon-like and
fluorine-like charge states of Fe are also extremely visible and there
is a very strong suggestion of emission from essentially the entire Fe
L series, from \Fexvii\ (neon-like) through to \Fexxiv\
(lithium-like).  Furthermore, the individual resonance,
intercombination (two lines, unresolved) and forbidden lines in the
\Ovii\ complex are resolved, as is some structure in the He-like
\Neix\ triple. These helium-like triples are discussed later. 

Though one cannot perform quantitative spectral fits to the fluxed
spectrum shown in Fig.~1, several conclusions can be drawn directly
from the figure. The strength of the \Fexvii\ lines, relative to the
K-shell lines, suggests strongly that collisional ionization is the
dominant mechanism for producing the soft X-ray emission. One can
infer temperatures of $\sim$0.3\,keV (\Ovii) to $\sim$1.5\,keV
(\Sixiii) or possibly higher. Significant photoelectric absorption
with implied column densities of a few 10$^{21}$ to
10$^{22}$\,cm$^{-2}$ is perhaps suggested by the weakness of the
longer wavelength lines. In fact, a comparison of Fig.~1 with the
fluxed RGS spectrum of the other famous nearby starburst galaxy,
NGC\,253 (Pietsch \etal\ 2001), is extremely instructive. Firstly, an
initial glance suggests that they are very similar, and in many
respects, they are. Concentrating on the differences, as regards the
weakness of the longer wavelength lines as compared with the shorter
wavelength ones, the relative deficit at long wavelengths in M82 may 
imply a larger absorbing column in M82 than in NGC\,253.
Furthermore, the greater strength of lines emanating from
higher-ionization-state ions relative to lower-ionization-state ions
(compare \Oviii\ to \Ovii, \Nex\ to \Nexi\ or \Fexviii\ to \Fexvii\ in
the two galaxies), together with a similar effect observed in lines
from the Fe$-$L series, implies a generally higher temperature in M82
than in NGC\,253.  As noted by Pietsch \etal\ (2001), the NGC\,253 RGS
spectrum, and by extension, the M82 RGS spectrum presented here,
appears very reminiscent of the spectrum of an intermediate-age
supernova remnant (SNR). This is not unexpected given the scenario of
a starburst nucleus and the interaction of the outflowing wind with
the surrounding cooler ISM gas. Lastly, it is worth noting the
difference in absolute flux levels $-$ M82 appears far brighter (by
about a factor of 4) than NGC\,253.

\section{Spectral Analysis}

X-ray spectroscopy in the RGS range is very well suited to studying
the mix in temperature and abundance of the relatively cool X-ray gas
in and around the M82 nuclear starburst. From what we already know we
can make some interesting conclusions. 

The fact that such a large range of Fe$-$L states exists
simultaneously, along with hydrogen- and helium-like emission lines
from N to Si, is indicative of the emission not being isothermal, and
point to emission from a multitemperature gas. Especially interesting
is the fact that there appears to exist perhaps all eight ionisation
states of the Fe$-$L series. This allows the possibility of using iron
as a type of thermometer; i.e. with ionisation potentials of between
500 and 2000\,eV, the distribution of electron temperatures in this
range can be addressed, and it is possible to uniquely determine the
temperature distribution. Highly detailed analysis using the Fe$-$L
series we defer to a later paper. It should be stated that the
determination of the temperature distribution using the H- and
He-like lines from the other different elements is plagued by
abundance uncertainties. Another point is that the emission features
from different ions appear to have different profiles and widths,
suggesting slight differences in the properties and/or relative
positions of the X-ray emitting gas within the M82 nuclear and
starburst areas. For example, the Mg and Ne lines appear rather thin,
whereas some of the Fe features have an appreciable width.

As the bright M82 X-ray core appears moderately resolved in the
RGS detectors, a rigorous analysis should also provide some
information as to the temperature and abundance structure. As we are
here dealing with merely a preliminary look at the M82 RGS data, and
as the interlinked spatial and thermal properties are likely to be
very complicated, we defer a rigourous discussion of such an analysis to a
later date. We have been
slightly unfortunate however, in that the RGS slit, lying parallel as it does
to the single long CCD gap of the EPIC-PN detector, samples the galaxy
in neither the major axis nor the minor axis direction, but rather at
an angle approximately bisecting the two. It would have been better
(and easier in analysis terms) to have the RGS sampling either along
the disk of the galaxy or along the direction of the starburst
outflow. We present an initial spatial analysis of the RGS data in the next 
section. 

In order to perform detailed spectral analysis, separate response files 
for each instrument and order (i.e. for each separate
spectrum created) were generated using {\em rgsrmfgen-0.45.3},
using, as is recommended, a large number (4000) of energy bins. These
responses were then attached to their relevant spectra and the spectra
were individually grouped into XSPEC-usable spectra with a minimum of
20 counts per bin, a bin typically spanning $\sim0.04$\,\AA. 

The four spectra (RGS 1 \& 2, orders 1 \& 2) were fitted
simultaneously using a variety of spectral models, forcing the model
parameters to be identical for each of the four spectral datasets, and
allowing only a small normalization difference between the four
components of the model(s) (to allow for small calibration
uncertainties between the instruments). In each model a redshift for
M82 of $z=0.0013$ was used. It is very useful to combine the data in
this manner as gaps in the data due to non-functioning CCDs
(specifically RGS1 CCD 7 and RGS2 CCD 4) can be filled with data from
the other instrument and order(s).

The fact that perhaps all the ionization states of the Fe$-$L series
exist indicates that the gas is clearly non-isothermal; i.e. gas over
a range of temperatures is contributing to the X-ray
emission. Physically realistic multi-temperature models were fit to
the data, the best (red.\,$\chi^{2}=1.35$, with 399 degrees of
freedom) being an XSPEC-{\em c6vmekl} model; a multi-temperature,
variable-abundance {\em mekal} model using a sixth-order Chebyshev
polynomial (Singh, White \& Drake 1996) for the differential emission
measure (DEM). The data and best-fitting model are shown in Fig.~2,
and the DEM is shown in Fig.~3. The XSPEC DEM shows emission from gas over a
range of temperatures, from around 0.2\,keV to 1.6\,keV, peaking at
around 0.7\,keV. Although the associated errors are large, super-solar 
abundances are obtained for Mg (3.3) and Si (7.1), moderate,
near-solar values are obtained for N (1.4), O (0.7) and Fe (0.9),
little Na (0.3) and essentially no Al is seen.

\begin{figure}
\vspace{6.0cm} 
\includegraphics{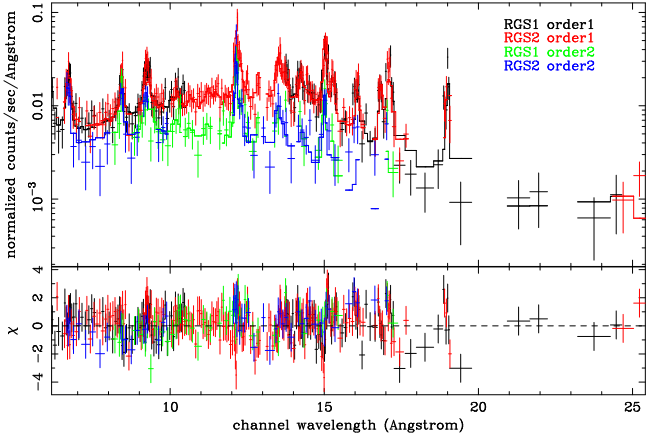} 
\caption{The RGS spectra from M82 together with the best-fitting
 multi-temperature,
variable-abundance {\em c6vmekl} model (see text). $\chi^{2}$ differences are
shown in the lower panel.}
\end{figure}

\begin{figure}
\vspace{5.0cm} 
\includegraphics{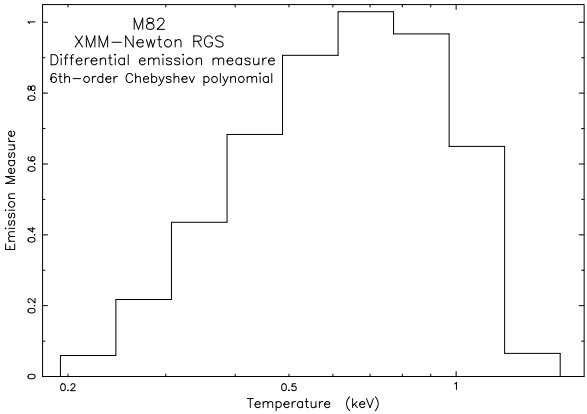} 
\caption{The differential emission measure from the best-fitting XSPEC{\em -c6vmekl} model 
to the M82 RGS spectra.}
\end{figure}

\subsection{Helium-like Triplets}

Within the helium-like \Ovii\ complex, the individual resonance ($r$),
intercombination ($i$, two lines, unresolved) and forbidden ($f$)
lines appear resolved, as is some structure in the He-like \Neix\
complex. A fit to the \Ovii\ complex, comprising of an absorbed
power-law, to represent the continuum, plus three fixed-energy
gaussians, to represent the triplet, was only conclusive in terms of
the $i$ line being significantly weaker than either of the other two
lines, and the $r$ line being stronger (perhaps by up to a factor 2)
then the $f$ line. Similar conclusions could be drawn by essentially
counting the photons appearing at the three energy positions in the
spectral data. Via a combination of these two methods, rough
approximations to the standard gas diagnostic ratios (Gabriel \&
Jordan 1969) of $R=f/i \gtsim 2.5$ and $G=(f+i)/r\approx 0.5-1$ could
be obtained. These values are consistent with that of collisionally
ionized gas, as expected in a hot gas starburst environment. The fact
that the $r$ line is stronger than the $f$ line is a further strong
indicator of hot gas in collisional ionization equilibrium (c.f. Fig.5
of Kinkhabwala \etal\ 2002). Little can be said as regards the \Neix\
triplet, due to strong contamination from \Fexix, \Fexx\ and \Fexix\
features at 13.5$-$13.8\,\AA.

\subsection{Line Widths and Shifts}

We attempted modelling the RGS spectra with a continuum plus a set of
spectral lines, the most successful model comprising a power-law to
represent the continuum with 11 gaussians representing the 11 most
prominent lines, specifically those of \Sixiii, \Mgxii, \Mgxi, \Nex,
\Neix, \Ovii\ and Fe$-$L lines at 0.863, 0.826, 0.775, 0.739 and
0.727\,keV. Fits were made using this model, both fixing the positions
of the lines as above, and also allowing for small redshifts in the
line positions (in order to perhaps detect some bulk motion of the
gas). Both fits resulted in very similar widths for the individual
lines, ranging from one or two 10's of eV; \Sixiii
(19.2$\pm$14.1\,eV), \Neix (10.6$\pm$6.7\,eV), Fe-L[0.863]
(22.1$\pm$5.5\,eV) to near-delta functions, e.g.\,the Mg lines (error
$\approx$\,5$-$10\,eV). Though a far better fit was obtained with the
redshifted-line model (red.\,$\chi^{2}=1.42$ c.f. 1.90), no trend was
seen within the errors as regards a large majority of the lines
showing the same redshift, and we cannot therefore attribute a bulk
velocity to the X-ray emitting gas on the basis of the X-ray lines
being systematically redshifted by the same amount.

\section{Spatial Analysis}

In looking at the line profiles in the dispersion direction, spatial
information can be extracted. This is easiest done when the line is
strong, non-complex (i.e. H-like), and is fairly far from any
neighbouring lines. The \Oviii\ line is a good example and shows
(Fig.~1) a larger tail of emission to lower wavelengths than to
higher. The RGS lies approximately east south-east to west north-west
(at a clockwise position angle to the vertical of
$\approx65^{\circ}$), and the \Oviii\ line profile is indicative of
more \Oviii\ emission lying to the east (and south-east) of the
nucleus, compared to the west (and north-west). Higher energy H-like
lines (\Neix, \Mgxii) exhibit more symmetric and thinner line
profiles, indicative of the emission in these higher energy lines as
being more centrally localised and being no more prevalent in one
direction around the nucleus than another.

As the RGS is a slitless, nearly stigmatic spectrometer, we can also
place some constraints on the spatial dependence of the various lines
by examining the emission line profiles in the {\em cross-dispersion} 
direction. In creating the cross-dispersion profile of the flux in a
particular emission line, events were selected (from, when available,
both the RGS1 and RGS2 data) with wavelength values within a certain
range (typically $\pm$0.25$-$0.5\,\AA) of the wavelength of the line
centre. Background events were extracted from an equal area (in
wavelength-cross-dispersion angle space) surrounding this initial line
extraction region. Once the line counts and background counts were
placed into 15\arcs\ cross-dispersion bins, the
background-subtracted cross-dispersion profile could be created. This
was done for each prominent line visible in the RGS spectrum. Four of
these profiles (from low to high energy) are shown in Fig.~4.

The RGS samples the galaxy emission (in
dispersion angle) at an angle approximately bisecting the major and
minor axes. The cross-dispersion axis therefore also bisects the major
and minor axes of M82 and lies such that negative distance values in
Fig.~4 correspond to areas to the south and south-west of the galaxy
nucleus (at a position angle clockwise from vertical of
$\approx155^{\circ}$). Positive values correspond to areas to the north
and north-east ($\approx-25^{\circ}$).

The \Oviii\ profile is the strongest and falls off less steeply to the
south (and SW) as to the north. A similar behaviour is seen in the
\Fexvii\ (15\AA) profile, and is indicative of there being more
emission in these lines to the south (and SW) of the nucleus as to the
north $-$ the nuclear emission appears more extended in these spectral
lines to the south than to the north. The higher energy \Nex\ profile
shows a similar behaviour also, but the profile is narrower (the FWHM
for \Oviii\ is about 110\arcs, whereas for \Nex, it is about
70\arcs). The emission from higher energy lines therefore appears more
centrally concentrated and not as extended. The low energy \Ovii\
feature, although weak, is interesting in that there appears to be an
enhancement at this energy to the north (and NE) of the nucleus.

\begin{figure}
\vspace{9.2cm} 
\includegraphics{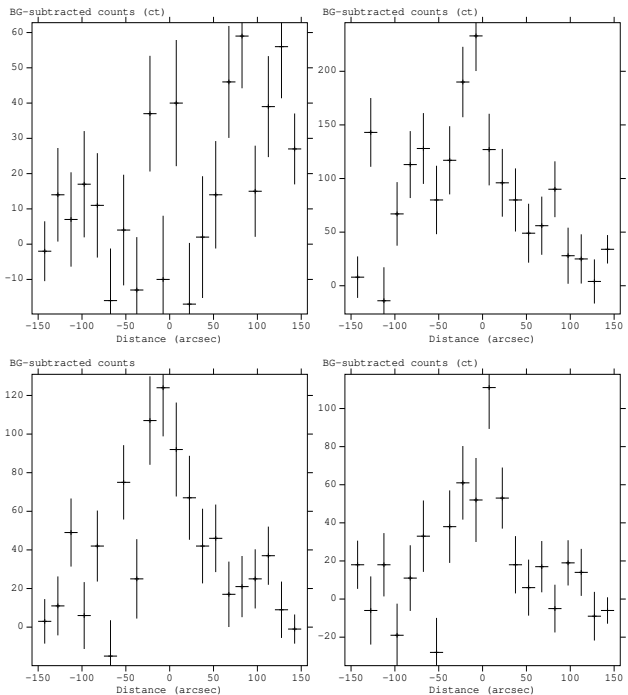} 
\caption{Profiles of background-subtracted counts against
cross-dispersion distance for the M82 RGS1 and RGS2 data in four
spectral lines; (top-left) \Ovii, (top-right) \Oviii, (bottom-left)
\Fexvii\ (15\AA), (bottom-right) \Nex. Note differences in Y-axis scales.}
\end{figure}

\section{Conclusions}

We have presented here the first high resolution spectra of
the M82 starburst taken with the {\em XMM-Newton} RGS
instrument. As M82 is the brightest starburst galaxy and {\em
XMM-Newton} is the satellite with the largest collecting area, these
spectra represent the best X-ray spectra to date of a starburst
galaxy. Initial spectral and spatial results have been presented
and these can be summarized as follows:

\begin{itemize}

\item The RGS spectrum is dominated by Ly$\alpha$ emission lines of
hydrogenic charge states of the abundant low Z elements (N, O, Ne, Mg,
Si). Helium-like charge states of the same elements are also seen in
emission, as is emission from the entire Fe$-$L series, from \Fexvii\
through to \Fexxiv.

\item The \Ovii\ complex is resolved into resonance, intercombination
and forbidden lines, and the line ratios are consistent with hot gas
in collisional ionization equilibrium. Significant structure is also
observed in the \Neix\ triple, though contamination from Fe exists.

\item The \Fexvii\ line strengths are consistent with emission from
gas in collisional ionization equilibrium. Gas over a range of
temperatures (0.3\,keV to 1.5\,keV, or higher) can be inferred from
the presence of the lines alone.

\item The M82 starburst appears generally hotter and brighter (by a
factor of $\sim$4) when compared with NGC~253. The weaker longer
wavelength lines may suggest a higher level of photoelectric
absorption in M82.

\item A multi-temperature, variable-abundance {\em mekal} model fits
the data well. The differential emission measure shows emission from
gas over a range of temperatures (from 0.2\,keV to 1.6\,keV, peaking at
around 0.7\,keV). High abundances are obtained for Mg and Si, while
near-solar values are obtained for N, O and Fe.

\item The \Oviii\ line profile suggests more \Oviii\ emission lying to
the east (and south-east) of the nucleus, compared to the west. The
more symmetric, thinner higher energy line profiles indicates their
emission as being more centrally localised and uniformly distributed.

\item Cross-dispersion line profiles suggest that emission from the
lower energy lines is more extended to the south (and SW).
Higher energy lines show similar behaviour, but the inferred extent is
reduced.

\end{itemize}

\section*{Acknowledgements} 

AMR and IRS acknowledge the support of PPARC funding, and thank the
referee for useful comments which have improved the paper
significantly.


\begin{thebibliography}{99}

\bibitem{}Cappi M., \etal, 1999, A\&A, 350, 777
\bibitem{}Bravo-Guerrero J., Read A.M., Stevens I.R., 2002, in prep.
\bibitem{}den Herder J.W., \etal, 2001, A\&A, 365, L7
\bibitem{}Devine D., Bally J., 1999, ApJ, 510, 197
\bibitem{}Freedman W.L., \etal, 1994, ApJ, 427, 628
\bibitem{}Gabriel A.H., Jordan C., 1969, MNRAS, 145, 241
\bibitem{}Heckman T.M., 2001, To appear in ``Extragalactic Gas at Low 
Redshift'', ed. J. Mulchaey and J. Stocke, ASP Conf. Series (astro-ph/0107438)
\bibitem{}Jansen F., \etal, 2001, A\&A, 365, L1
\bibitem{}Kaaret P., \etal, 2001, MNRAS, 321, L29
\bibitem{}Kinkhabwala A., \etal, 2002, ApJ, in press (astro-ph/0203290)
\bibitem{}Lehnert M.D., Heckman T.M., Weaver K.A., 1999, ApJ,  523, 575
\bibitem{}Pietsch W., \etal, 2001, A\&A, 365, L174
\bibitem{}Read A.M., Ponman T.J., Strickland D.K., 1997, MNRAS, 286, 626 
\bibitem{}Rieke G.H., Loken K., Rieke M.J., Tamblyn P., 1993, ApJ, 412, 99
\bibitem{}Shopbell P.L., Bland-Hawthorn J., 1998, ApJ, 493, 129
\bibitem{}Singh K.P., White N.E., Drake S.A., 1996, ApJ, 456, 766
\bibitem{}Strickland D.K., Ponman T.J., Stevens I.R., 1997, A\&A, 320, 378\\

\end{thebibliography}
\end{document}